\documentclass[aps,tightenlines,
english,preprintnumbers,amsmath,amssymb,nofootinbib,superscriptaddress]{revtex4-1}

\usepackage{bbm}
\usepackage{bm,soul}
\usepackage{amsmath}
\usepackage{mathrsfs}
\usepackage{slashed}
\usepackage{keyval,amsfonts,slashed,bm}
\usepackage{graphicx,subfigure,textcomp}
\usepackage[colorlinks=true,linktocpage=true,linkcolor=blue,citecolor=blue]{hyperref}

\def\nn{\nonumber\\}

\def\lb{\left(}
\def\rb{\right)}
\def\hmu{\hat\mu}
\def\L{\ln\frac{\hat\Lambda}{2}}

\def\Lg{\ln\frac{\hat\Lambda_g}{2}}
\def\Za{\frac{\zeta'(-1)}{\zeta(-1)}}
\def\Zc{\frac{\zeta'(-3)}{\zeta(-3)}}
\def\bqa{\begin{eqnarray}}
\def\eqa{\end{eqnarray}}

\def\[{\left[}
\def\]{\right]}  
\def\({\left(}
\def\){\right)}  
\def\hmD{\hat m_D}

\begin{document}

\author{Jens O. Andersen}
\affiliation{Department of Physics, Norwegian University of Science and  
Technology, N-7491 Trondheim, Norway}
\author{Najmul Haque}
\affiliation{Department of Physics, Kent State University, Kent, Ohio 44242, 
United States}
\author{Munshi G. Mustafa}
\affiliation{Theory Division, Saha Institute of Nuclear Physics, 1/AF 
Bidhannagar, Kolkata-700064, India}
\author{Michael Strickland}
\affiliation{Department of Physics, Kent State University, Kent, Ohio 44242, 
United States}


\title{Three-loop hard-thermal-loop perturbation theory thermodynamics at finite temperature and finite baryonic and isospin chemical potential
}


\begin{abstract}
In a previous paper (JHEP {\bf 05} (2014) 27), we calculated the 
three-loop thermodynamic potential of QCD at finite
temperature $T$ and quark chemical potentials $\mu_q$
using the hard-thermal-loop perturbation theory 
(HTLpt) reorganization of finite temperature and density QCD. 
The result allows us to study the thermodynamics of QCD at finite
temperature and finite baryon, strangeness, and  
isospin chemical potentials $\mu_B$, $\mu_S$, and $\mu_I$.
We calculate the pressure at nonzero $\mu_B$ and $\mu_I$ with $\mu_S=0$,
and the energy density, entropy density, 
the trace anomaly, and the speed of sound at 
nonzero $\mu_I$ with $\mu_B=\mu_S=0$.
The second and fourth-order isospin susceptibilities
are calculated at $\mu_B=\mu_S=\mu_I=0$.
Our results can be directly compared to lattice QCD without Taylor
expansions around $\mu_q=0$ since QCD has no sign problem at 
$\mu_B=\mu_S=0$ and finite isospin chemical potential $\mu_I$.
\end{abstract}
\maketitle


\section{Introduction}
\label{intro}
Quantum chromodynamics (QCD) in extreme conditions such as high
temperature and high density has been a very active area of research
for more than two decades.
The interest in QCD at finite temperature has largely been spurred
by the experimental programs in heavy-ion collisions at the 
Relativistic Heavy Ion Collider (RHIC) in Brookhaven
and the Large Hadron Collider
(LHC) at CERN.
One of the goals of these programs is the creation and study of the
quark-gluon plasma - the deconfined phase of QCD.
The equation of state (EoS) of QCD is essential to the phenomenology
of the quark-gluon plasma. 
Lattice gauge theory
provides a first-principle method to calculate the thermodynamic functions
of QCD at finite temperature and zero baryon chemical potential $\mu_B$.
However, at finite $\mu_B$, QCD suffers from the so called sign problem,
namely that the fermion determinant is complex. This prevents 
one from using standard lattice techniques involving importance sampling to
calculate the partition function of QCD. One way to circumvent this problem,
at least for small baryon chemical potentials, is to make a Taylor
expansion of the thermodynamic functions around $\mu_B$. This requires
the calculation of the quark-number susceptibilities evaluated at 
zero quark chemical potentials, $\mu_q=0$.

Perturbative QCD offers an alternative to lattice gauge theory
for the calculations of thermodynamic functions in the 
deconfined phase. Invoking asymptotic freedom, one might expect 
that perturbation theory works
at sufficiently high temperatures. However, one does
not know a priori how large $T$ must be in order to obtain 
a sufficiently good approximation. Using the weak-coupling
expansion in the strong coupling constant $g$, 
the calculation of the thermodynamic functions has been pushed
to order $g^6\log g$ both at zero~\cite{kajantie}
and finite chemical potential~\cite{vuorinen1,vuorinen2,ipp}.
However, a strict perturbative expansion in $g$ does not converge
at temperatures relevant for the heavy-ion collision experiments.
It turns out that the convergence is very poor unless the temperature
is many orders of magnitude larger than the critical temperature $T_c$
for the deconfinement transition.
The source of the poor convergence is the contributions to the thermodynamic
functions coming from soft momenta of order $gT$.
The poor convergence of the weak-coupling expansion suggests that one
needs to reorganize the perturbative series of thermal QCD.
For scalar theories, screened perturbation theory 
(SPT) has been applied 
successfully up to four loops~\cite{spt1,spt2,spt3,spt4}. 
SPT is in part inspired by variational perturbation 
theory~\cite{vpt1,vpt2,vpt3,vpt4,vpt5,vpt6},
see also~\cite{kneur} for a renormalization-group improved
reorganization of the perturbative series.
In the case of gauge theories, using a local mass term for the gluons breaks
gauge invariance and one needs to generalize SPT.
Hard-thermal-loop perturbation theory (HTLpt) represents such a 
generalization and was developed over a decade ago~\cite{andersen1}.
Since its invention, HTLpt has been used to calculate thermodynamic
functions through three loops at 
zero chemical 
potential~\cite{3loopglue1,3loopglue2,3loopqcd1,3loopqcd2,3loopqcd3} 
as well as finite chemical potential~\cite{najmul3,complete}.
Depending on the thermodynamic function
at hand, the agreement between lattice simulations
and the results from HTLpt is very good down to temperatures of approximately
$T\simeq 250$ MeV. 
Application of some HTL-motivated approaches can be found
in Refs.~\cite{purnendu1,purnendu2,purnendu3,najmul11,najmul12,
najmul13,blaizot1,blaizotm1,blaizotm2,blaizotm3,blaizot2,blaizot3}.

While three-color 
QCD at finite baryon chemical potential has a sign problem, there
are a number of other cases where the sign problem is absent. This includes
QCD in a strong magnetic field $B$, two-color QCD at finite baryon chemical
potential $\mu_B$~\cite{twocolor1,twocolor2}, 
and three-color QCD at finite isospin chemical potential 
$\mu_I$\cite{alfiso}. 
In this paper, we will focus on three-color QCD at finite isospin density.
There are a few papers on lattice QCD with finite isospin chemical 
potential~\cite{latt0,latt1,latt2,allton,ejiri,forc,detmold,endiso}, 
however, these mostly focus on the phase
transitions themselves and not on the deconfined phase:
In addition to the deconfinement transition, 
there is an additional transition 
to a Bose condensate
of pions at sufficiently low temperature $T$ and 
sufficiently large 
isospin chemical potential $\mu_I$~\cite{stepson}.
For $T=0$, the critical chemical potential for pion condensation is
$\mu_I^c=m_{\pi}$. Moreover, the results of~\cite{forc} seem to indicate
that the first-order deconfinement transition at zero isospin density
turns into a crossover at $\mu_I/T\simeq2.5$.
At sufficiently low temperature and high isospin chemical potential, i.e.
around the phase boundary, HTLpt is unreliable.
Thus, at this point in time, we cannot compare our HTLpt predictions with
lattice Monte Carlo at finite $\mu_I$. 
Therefore, our results should be
considered as predictions which can be checked by future
lattice simulations.
This is in contrast to three-color QCD at $\mu_B=0$, where there is a plethora
of lattice results~\cite{borsanyi1,borsanyi2,Borsanyi:2012uq,borsanyi3,borsanyi4,borsanyi5,sayantan,bnlb0,bnlb1,bnlb2,bnlb3,milc,hotqcd1,hotqcd2,peter_review,hotqcd15,dingrev,bellwied,ding} on the thermodynamics of the
deconfined phase.

The paper is organized as follows. 
In section~\ref{densi}, we briefly discuss finite chemical potentials
and the sign problem of QCD.
In section~\ref{htlpt}, we review hard-thermal-loop perturbation theory
and the HTLpt thermodynamic potential through next-next-to-leading order
(NNLO).
In section~\ref{numerical}, we
present and discuss our numerical results for the thermodynamic functions.
In section~\ref{conclude}, we summarize and conclude.

\section{Particle densities, chemical potentials, and the sign problem in QCD}
\label{densi}
In massless QCD with $N_f$ flavors there are $N_f^2$ conserved
charges which corresponds to the number of generators of the
group $SU(N_f)\times U(1)$.
For each conserved charge $Q_i$, we can introduce a nonzero chemical potential
$\mu_i$. However, it is possible to specify the expectation values of
different charges simultaneously, only if they commute.  For $N_f=2$
and $N_f=3$, this implies that we can introduce two and three
independent chemical potentials, respectively.
These can conveniently be chosen as the quark chemical potentials $\mu_q$,
which corresponds to the separate conservation of the number of
$u$, $d$, and $s$ quarks.
However, any other independent linear combination of $\mu_q$
is equivalent and it is customary to introduce chemical potentials
for baryon number $n_B$, isospin $n_I$, and strangeness $n_S$.

After having introduced the chemical potentials in the Lagrangian,
the partition function as well as all thermodynamic quantities are 
functions of the temperature and the chemical potentials.
For example, the corresponding charge densities $n_i$ are
given by
\bqa
n_i&=&-{\partial{\cal F}\over\partial\mu_i}\;,
\label{charges}
\eqa
where $\cal F$ is the free energy density.

The baryon, isospin, and strangeness densities $n_B$, $n_I$, and $n_S$
can be expressed in terms of the quark densities $n_f$ as
\bqa
\label{rel1}
n_B&=&{1\over3}(n_u+n_d+n_s)\;,\\
\label{rel2}
n_I&=&
n_u-n_d\;, \\
n_S&=&-n_s\;.
\label{rel3}
\eqa
Eqs.~(\ref{rel1})--(\ref{rel3}) can be used to derive relations
between the corresponding chemical potentials $\mu_B$, $\mu_I$, and $\mu_S$
and the quark chemical potentials $\mu_q$. 
Eqs.~(\ref{charges}) and~(\ref{rel2}) give
\bqa\nonumber
n_I&=&-{\partial{\cal F}\over\partial\mu_I}\\ \nonumber
&=&-
\left({\partial{\cal F}\over\partial\mu_u}
-{\partial{\cal F}\over\partial\mu_d}\right)
\\
&=&-\left(
{\partial\mu_u\over\partial\mu_I}
{\partial{\cal F}\over\partial\mu_u}
+{\partial\mu_d\over\partial\mu_I}
{\partial{\cal F}\over\partial\mu_d}\right)\:.
\label{relmu}
\eqa
Comparing second and third line in Eq. (\ref{relmu}),
we infer that
\bqa
{\partial\mu_u\over\partial\mu_I}
=-{\partial\mu_d\over\partial\mu_I}
=1
\;.
\eqa
In the same manner, one can show that 
${\partial\mu_u\over\partial\mu_B}={\partial\mu_d\over\partial\mu_B}
={\partial\mu_s\over\partial\mu_B}={1\over3}$,
${\partial\mu_u\over\partial\mu_S}={\partial\mu_d\over\partial\mu_S}=0$,
and ${\partial\mu_s\over\partial\mu_S}=-1$.
This gives the following
relations between the chemical potentials $\mu_B$, $\mu_I$,
and $\mu_S$ and the quark chemical potentials $\mu_q$
\bqa
\label{rel11}
\mu_u&=&{1\over3}\mu_B+
\mu_I\;,\\
\mu_d&=&{1\over3}\mu_B-
\label{rel22}
\mu_I\;,\\
\mu_s&=&{1\over3}\mu_B-\mu_S\;.
\label{rel33}
\eqa
In the chiral (Weyl) representation, 
we can write the Dirac operator 
$(D\!\!\!\!/+m-\mu_q\gamma_0)$
for three flavors as
\begin{widetext}
\bqa\nonumber
\small
\left(
\begin{array}{cccccc}
m&iX-{1\over3}\mu_B-\mu_I&0&0&0&0\\
iX^{\dagger}-{1\over3}\mu_B-\mu_I&m&0&0&0&0 \\
0&0&m&iX-{1\over3}\mu_B+\mu_I&0&0\\
0&0&iX^{\dagger}-{1\over3}\mu_B+\mu_I&m&0&0 \\
0&0&0&0&m&iX-{1\over3}\mu_B+\mu_S\\
0&0&0&0&iX^{\dagger}-{1\over3}\mu_B+\mu_S&m \\
\end{array}
\right)\;,
\\
\eqa
where 
$iX=D_0+i{\boldsymbol \sigma}\cdot{\bf D}$
. 
The fermion determinant then becomes
\bqa\nonumber
\det(D\!\!\!\!/+m-\mu_q\gamma_0)&=&
\det\left[\left(X^{\dagger}+\mbox{$1\over3$}i\mu_B+i\mu_I)(X+\mbox{$1\over3$}
i\mu_B+i\mu_I\right)+m^2\right]
\\&& \nonumber
\times\det
\left[\left(X^{\dagger}+\mbox{$1\over3$}i\mu_B-i\mu_I)(X+\mbox{$1\over3$}i\mu_B-i\mu_I\right)+m^2\right]
\\&& \times\det
\left[
\left(X^{\dagger}+\mbox{$1\over3$}i\mu_B-i\mu_S)(X+\mbox{$1\over3$}i\mu_B-i\mu_S\right)
+m^2\right]\;.
\label{det}
\eqa
The terms proportional to $\mu_B$ and $\mu_S$
appear in the same way in combination with $X^{\dagger}$
and $X$. Consequently, the 
fermion determinant is real only for $\mu_B=\mu_S=0$.
Using Eqs.~(\ref{rel11})--(\ref{rel33}), this
yields the constraints
\bqa
\mu_u+\mu_d&=&0\;,\\
\mu_s&=&0\;.
\eqa
Given the two constraints, there is only one independent 
chemical potential, for example, the isospin chemical potential 
$\mu_I={1\over2}(\mu_u-\mu_d)$.
The fermion determinant
reduces to
\bqa
\det(D\!\!\!\!/+m-\mu_q\gamma_0)&=&
\det\left[\left(X^{\dagger}+i\mu_I)(X+i\mu_I\right)+m^2\right]
\det
\left[\left(X^{\dagger}-i\mu_I)(X-i\mu_I\right)+m^2\right]
\det\left[X^{\dagger}X+m^2\right]\;.
\eqa
\end{widetext}
We conclude that the fermion determinant is real even
for nonzero isospin chemical
potential and this proves that there is no sign problem for $\mu_I\neq0$.

\section{Hard-thermal-loop perturbation theory}
\label{htlpt}
In this section, we briefly review hard-thermal-loop perturbation theory.
For a detailed discussion, see for example Ref.~\cite{complete}.
Hard-thermal-loop perturbation theory is a reorganization of 
perturbation theory for thermal QCD. The HTLpt 
Lagrangian density is written as 
\bqa
 {\cal L}=
\left.({\cal L}_{\rm QCD}+{\cal L}_{\rm HTL})
\right|_{g\rightarrow\sqrt{\delta}g}+\Delta{\cal L}_{\rm HTL} \, , 
\label{total_lag} 
\eqa
where the HTL improvement term is~\cite{lagrangian} 
\begin{widetext}
\bqa
 {\cal L}_{\rm HTL}=(1-\delta)i m_q^2
\bar\psi\gamma^\mu\left\langle\frac{y_\mu}{y\cdot\! D}\right\rangle_{\!\hat{\bf y}}
\psi-\frac{1}{2}(1-\delta)
 m_D^2 {\rm Tr}\lb G_{\mu\alpha}\left\langle\frac{y^\alpha y_\beta}{(y\cdot\! D)^2}
\right\rangle_{\!\hat{\bf y}} G^{\mu\beta}\rb \, ,
\label{htl_lag}
\eqa
\end{widetext}
and $\Delta{\cal L}_{\rm HTL}$ contains additional HTLpt counterterms.
Here $y^\mu = (1, {\bf\hat{y}})$ is a light-like four-vector 
with ${\bf\hat{y}}$ being a three-dimensional unit vector and the angular 
bracket indicates an average over the direction of ${\bf\hat{y}}$. The two 
parameters $m_D$ and $m_q$ can be identified with the Debye screening mass and 
the thermal quark mass, respectively, and account for screening effects.  
HTLpt is defined by treating $\delta$ as a formal expansion parameter. 
The HTLpt Lagrangian (\ref{total_lag}) reduces to the QCD Lagrangian 
if we set $\delta=1$.  Physical observables are calculated in 
HTLpt by expanding in powers of $\delta$, truncating at some specified order, 
and setting $\delta = 1$ in the end.  This defines a reorganization of the 
perturbative series in which the effects of $m_D^2$ and $m_q^2$ terms in 
(\ref{htl_lag}) are included to leading order but then systematically 
subtracted out at higher orders.
Note that HTLpt is gauge invariant order-by-order in the $\delta$ expansion 
and, consequently, the results obtained are independent of the gauge-fixing 
parameter $\xi$ (in the class of covariant gauges we are using).
To zeroth order in $\delta$, HTLpt describes a gas of massive 
gluonic and quark quasiparticles. Thus, HTLpt 
systematically shifts the perturbative expansion 
from being around an ideal gas of massless particles to being around a gas of 
massive quasiparticles which are the appropriate physical degrees of freedom 
at high temperature and/or chemical potential.

Higher orders in $\delta$ describe the interaction among these quasiparticles
and involve standard QCD Feynman diagrams as well new diagrams generated
by the HTL improvement term.
If the expansion in $\delta$ could be calculated to all orders, the final 
result would not depend on $m_D$ and $m_q$ when we set $\delta=1$. However, 
any truncation of the expansion in $\delta$ produces results that depend on 
$m_D$ and $m_q$. As a consequence, a prescription is required to determine 
$m_D$ and $m_q$ as a function of $T$, $\mu_q$, and $\alpha_s$. Several 
prescriptions were discussed in~\cite{3loopqcd2} at zero chemical 
potential and generalized to finite chemical potential in~\cite{complete}. 
We return to this issue below.

\subsection{NNLO HTLpt thermodynamic potential}
The QCD free energy to three-loop order in HTLpt for the case
that each quark $f$ has a separate quark chemical potential $\mu_f$
was calculated in \cite{complete}. 
The final result is
\begin{widetext}
\begin{eqnarray}
\frac{\Omega_{\rm NNLO}}{\Omega_0}
&=& \frac{7}{4}\frac{d_F}{d_A}\frac{1}{N_f}
\sum\limits_f\lb1+\frac{120}{7}\hmu_f^2+\frac{240}{7}\hmu_f^4\rb
    -\frac{s_F\alpha_s}{\pi}\frac{1}{N_f}\sum\limits_f\bigg[\frac{5}{8}
\left(1+12\hat\mu_f^2\right)\left(5+12\hat\mu_f^2\right)
    \nn
    &&-\frac{15}{2}\left(1+12\hat\mu_f^2\right)\hat m_D-\frac{15}{2}
\bigg(2\ln{\frac{\hat\Lambda}{2}-1
   -\aleph(z_f)}\Big)\hat m_D^3
      +90\hat m_q^2 \hat m_D\bigg]
\nn
&&+ \frac{s_{2F}}{N_f}\left(\frac{\alpha_s}{\pi}\right)^2\sum\limits_f
\bigg[\frac{15}{64}\bigg\{35-32\lb1-12\hmu_f^2\rb\frac{\zeta'(-1)}
      {\zeta(-1)}+472 \hat\mu_f^2+1328  \hat\mu_f^4\nn
      &&+ 64\Big(-36i\hat\mu_f\aleph(2,z_f)+
6(1+8\hat\mu_f^2)\aleph(1,z_f)+3i\hat\mu_f(1+4\hat\mu_f^2)\aleph(0,z_f)\Big)
\bigg\}\nn
      &&- \frac{45}{2}\hat m_D\left(1+12\hat\mu_f^2\right)\bigg] \nn
&&+ \left(\frac{s_F\alpha_s}{\pi}\right)^2
      \frac{1}{N_f}\sum\limits_{f}\frac{5}{16}\Bigg[96\left(1+12\hat\mu_f^2
\right)\frac{\hat m_q^2}{\hat m_D}
     +\frac{4}{3}\lb1+12\hmu_f^2\rb\lb5+12\hat\mu_f^2\rb
      \ln\frac{\hat{\Lambda}}{2}\nn
    && +\frac{1}{3}+4\gamma_E+8(7+12\gamma_E)\hat\mu_f^2+112\hmu_f^4-
\frac{64}{15}\frac{\zeta^{\prime}(-3)}{\zeta(-3)}-
   \frac{32}{3}(1+12\hat\mu_f^2)\frac{\zeta^{\prime}(-1)}{\zeta(-1)}\nn
   &&-    96\Big\{8\aleph(3,z_f)+12i\hat\mu_f\aleph(2,z_f)-2(1+2\hat\mu_f^2)
\aleph(1,z_f)-i\hat\mu_f\aleph(0,z_f)\Big\}\Bigg] \nn
&&+ \left(\frac{s_F\alpha_s}{\pi}\right)^2
      \frac{1}{N_f^2}\sum\limits_{f,g}\left[\frac{5}{4\hat m_D}
\left(1+12\hat\mu_f^2\right)\left(1+12\hat\mu_g^2\right)
     +90\Bigg\{ 2\left(1 +\gamma_E\right)\hat\mu_f^2\hat\mu_g^2
      \right.\nn
        &&-\Big\{\aleph(3,z_f+z_g)+\aleph(3,z_f+z_g^*)+ 4i\hat\mu_f
\left[\aleph(2,z_f+z_g)+\aleph(2,z_f+z_g^*)\right]-4\hat\mu_g^2\aleph(1,z_f)\nn
       &&
       -(\hat\mu_f+\hat\mu_g)^2\aleph(1,z_f+z_g)- (\hat\mu_f-\hat\mu_g)^2
\aleph(1,z_f+z_g^*)-4i\hat\mu_f\hat\mu_g^2\aleph(0,z_f)\Big\}\Bigg\}\nn
       &&-\left.\frac{15}{2}\lb1+12\hat\mu_f^2\rb\lb2\L-1-\aleph(z_g)\rb  
\hat m_D\right]
\nn
&&+ \left(\frac{c_A\alpha_s}{3\pi}\right)\left(\frac{s_F\alpha_s}{\pi N_f}
\right)\sum\limits_f\Bigg[\frac{15}{2\hat m_D}\lb1+12\hmu_f^2\rb
     -\frac{235}{16}\Bigg\{\bigg(1+\frac{792}{47}\hat\mu_f^2+\frac{1584}{47}
\hat\mu_f^4\bigg)\ln\frac{\hat\Lambda}{2}
     \nonumber\\
    &&-\frac{144}{47}\lb1+12\hmu_f^2\rb\ln\hat m_D+\frac{319}{940}
\left(1+\frac{2040}{319}\hat\mu_f^2+\frac{38640}{319}\hat\mu_f^4\right)
   -\frac{24 \gamma_E }{47}\lb1+12\hat\mu_f^2\rb
\nonumber\\
    &&
   -\frac{44}{47}\lb1+\frac{156}{11}\hmu_f^2\rb\frac{\zeta'(-1)}{\zeta(-1)}
    -\frac{268}{235}\frac{\zeta'(-3)}{\zeta(-3)}
   -\frac{72}{47}\Big[4i\hat\mu_f\aleph(0,z_f)+\left(5-92\hat\mu_f^2\right)
\aleph(1,z_f)
    \nonumber\\
    &&+144i\hmu_f\aleph(2,z_f)
   +52\aleph(3,z_f)\Big]\Bigg\}+90\frac{\hat m_q^2}{\hat m_D}+\frac{315}{4}
\Bigg\{\lb1+\frac{132}{7}\hmu_f^2\rb\L
\nonumber\\
   &&+\frac{11}{7}\lb1+12\hmu_f^2\rb\gamma_E+\frac{9}{14}\lb1+\frac{132}{9}
\hmu_f^2\rb
+\frac{2}{7}\aleph(z_f)\Bigg\}\hat m_D 
\Bigg]
+ \frac{\Omega_{\rm NNLO}^{\rm YM}}{\Omega_0} \, ,
\label{finalomega}
\end{eqnarray}
where $\Omega_0=- \frac{d_A\pi^2 T^4}{45}$, 
$\hat{\mu}_f=\mu_f/2\pi T$, 
$\hat{\Lambda}=\Lambda/2\pi T$, 
and $\hat{m}_D=m_D/2\pi T$.
The QCD Casimir numbers are
$c_A=N_c$, $d_A=N^2_c-1$, $s_F=N_f/2$, $d_F=N_cN_f$, and
$s_{2F}=C_Fs_F$ with $C_F= (N^2_c-1)/2N_c$.
The sums over $f$ and $g$ include all quark flavors, 
$z_f = 1/2 - i \hat{\mu}_f$, and $\Omega_{\rm NNLO}^{\rm YM}$ is the pure-glue 
contribution  
\bqa
\frac{\Omega_{\rm NNLO}^{\rm YM}}{\Omega_0} &=& 1-\frac{15}{4}\hat m_D^3+
\frac{c_A\alpha_s}{3\pi}\Bigg[-\frac{15}{4}
+\frac{45}{2}\hat m_D-\frac{135}{2}\hat m_D^2-\frac{495}{4}\lb\Lg+\frac{5}{22}
+\gamma_E\rb  \hat m_D^3 \Bigg]
\nn
&+&\lb\frac{c_A\alpha_s}{3\pi}\rb^2\Bigg[\frac{45}{4\hat m_D}-\frac{165}{8}\lb
\Lg-\frac{72}{11}\ln\hat m_D-\frac{84}{55}-\frac{6}{11}
\gamma_E-\frac{74}{11}\Za+\frac{19}{11}\Zc\rb
\nn
&+&\frac{1485}{4}\lb\Lg-\frac{79}{44}+\gamma_E+\ln2-\frac{\pi^2}{11}\rb\hat m_D
\Bigg] \, .
\label{ymomega1}
\eqa
\end{widetext}
In Eq.~(\ref{finalomega}), the functions $\aleph(z)$ and $\aleph(n,z)$
appear. These are defined as
\bqa
\aleph(z)&=&\Psi(z)+\Psi(z^*)\;,
\\ 
\aleph(n,z)&=&\zeta^{\prime}(-n,z)+(-1)^{n+1}\zeta^{\prime}(-n,z^*)\;,
\eqa
where 
\bqa
\zeta^{\prime}(x,y)&=&\partial_x\zeta(x,y)\;,
\\
\Psi(z)&=&{\Gamma^{\prime}(z)\over\Gamma(z)}\;.
\eqa
Here $\zeta(x,y)$ is the Riemann zeta function and $\Gamma(z)$
is the digamma function.

\subsection{Mass prescription}
\label{pres}
In order to complete a calculation in HTLpt, we must 
have a prescription for the mass parameters $m_D$ and $m_q$
appearing in the HTL Lagrangian. A variational prescription seems natural,
i.e. one looks for solutions of
\bqa
{\partial\over\partial m_D}\Omega(T,\alpha_s,m_D,m_q,\mu_q,\delta=1)
&=&0\;,
\\
{\partial\over\partial m_q}\Omega(T,\alpha_s,m_D,m_q,\mu_q,\delta=1)
&=&0\;.
\eqa
However, in some case the resulting gap equations only have complex solutions
and one must look for other prescriptions.
Inspired by dimensions reduction, one equates the
mass parameter $m_D$ with the mass parameter of three-dimensional
Electric QCD (EQCD) in~\cite{braatennieto2}.
This mass can be interpreted as the contribution to the Debye mass from the 
hard scale $T$and is
well defined and gauge invariant order-by-order in perturbation theory.
This prescription was used in Ref.~\cite{complete}
and will be
used  in the remainder of the paper as well.
Originally, the two-loop perturbative mass was 
calculated in Ref.~\cite{braatennieto2} for zero chemical 
potential, however, Vuorinen has generalized it to finite chemical potential.  
The resulting expression for $\hat m_D^2$ is~\cite{vuorinen1,vuorinen2}
\begin{widetext}
\begin{eqnarray}
\hat m_D^2&=&\frac{\alpha_s}{3\pi} \Biggl\{c_A
+\frac{c_A^2\alpha_s}{12\pi}\lb5+22\gamma_E+22\Lg\rb +
\frac{1}{N_f} \sum\limits_{f}
\Biggl[ s_F\lb1+12\hmu_f^2\rb
\nonumber\\
  &&+\frac{c_As_F\alpha_s}{12\pi}\lb\lb9+132\hmu_f^2\rb+22\lb1+12\hmu_f^2\rb\gamma_E+2\lb7+132\hmu_f^2\rb\L+4\aleph(z_f)\rb
\nonumber\\  
&&+\frac{s_F^2\alpha_s}{3\pi}\lb1+12\hmu_f^2\rb\lb1-2\L+\aleph(z_f)\rb
 -\frac{3}{2}\frac{s_{2F}\alpha_s}{\pi}\lb1+12\hmu_f^2\rb \Biggr] \Biggr\} \, .
\label{debyev}
\end{eqnarray}
\end{widetext}
The effect of the in-medium quark mass parameter $m_q$ in thermodynamic 
functions is small and following
Ref.~\cite{3loopqcd2}, we take $m_q=0$.
  


\section{Numerical results}
\label{numerical}
In this section, we present our results for the NNLO HTLpt thermodynamic
functions at finite temperature $T$ and isospin chemical potential
$\mu_I$, and $\mu_B=\mu_S=0$. We emphasize that all thermodynamic
functions can be calculated for nonzero values of the three
independent chemical potentials.

\subsection{Running coupling and scales}
In Ref.~\cite{3loopqcd2}, we showed that the renormalization of the
three-loop HTLpt free energy is consistent with the standard
one-loop running of the strong coupling constant~\cite{run1,run2}.
Using a one-loop running is therefore self-consistent
and will be used in the remainder of this 
paper.\footnote{In our previous paper~\cite{complete}, we used one-loop 
running as
well as three-loop running to gauge the sensitivity of our results.
Generally, our three-loop HTLpt predictions
were rather insensitive to whether we used 
one-loop or three-loop running.}
In this case, the running coupling 
$\alpha_s(\Lambda)$ is given by 
%
\bqa
\alpha_s(\Lambda)&=&\frac{1}{b_0 t}\;,
\eqa
with $t = \ln(\Lambda^2/\Lambda_{\overline{\rm MS}}^2)$ and 
$b_0=(11c_A-2N_f)/12\pi$.
%
We fix the scale $\Lambda_{\overline{\rm MS}}$ by requiring that 
$\alpha_s({\rm 1.5\;GeV}) = 0.326$ which is obtained from
independent lattice measurements~\cite{latticealpha}.  
For one-loop running, this procedure gives 
$\Lambda_{\overline{\rm MS}} = 176$ MeV.
MeV.

For the renormalization scale we use separate scales, $\Lambda_g$ and 
$\Lambda_q$, for purely-gluonic and fermionic graphs, respectively.  We take 
the central values of these renormalization scales to be $\Lambda_g = 2\pi T$ 
and $\Lambda=\Lambda_q=2\pi \sqrt{T^2+(\mu_B^2+2\mu_I^2)/(N_f\pi^2)}$.  
In all plots, the 
thick 
lines indicate the result obtained using these central values and the 
light-blue band indicates the variation of the result under variation of both 
of these scales by a factor of two, e.g. $\pi T \leq \Lambda _g \leq 4 \pi T$.  
For all numerical results below we use $c_A = N_c=3$ and $N_f=3$.


Since our final result for the thermodynamic potential 
(\ref{finalomega}) and the thermodynamic functions that are derived
from it, are expansions in  $m_D/T$ and $m_q/T$, we cannot push our results to
very high values of $\mu_I$; the Debye mass in Eq.(\ref{debyev})
depends on the quark chemical potentials $\mu_f$.
An estimate for the reliability of HTLpt is that $m_D\simeq g T$.
If $T<\sqrt{3}\mu_f/\pi$, the $\mu_f$-dependent term of $m_D$ just starts to 
dominate over
the $T$-dependent term. Thus we consider $\mu_f\lesssim \pi T$ as reasonable. 
For temperatures
down to $150$ MeV, we decide to err on the safe side and use 
$\mu_I$ no larger than $400$ MeV.

\begin{widetext}
\subsection{Pressure}
The pressure of the quark-gluon plasma can be obtained directly from the 
thermodynamic potential~(\ref{finalomega})
\bqa
{\cal P}(T,\Lambda,\mu_u,\mu_d,\mu_s)
&=&
-\Omega(T,\Lambda,\mu_u,\mu_d,\mu_s)\;,
\label{thermo}
\eqa
where $\Lambda$ includes both $\Lambda_g$ and $\Lambda_q$.
The pressure can be
obtained using our general
expression Eq.~(\ref{finalomega}) for nonzero values of $\mu_B$ and $\mu_I$
and for $\mu_S=0$ using the Eqs.~(\ref{rel11})-(\ref{rel33}). For simplicity, we
are
presenting here the NNLO HTL pressure only at nonzero value of $\mu_I$ and for 
$\mu_B=\mu_S=0$ as

\begin{eqnarray}
\frac{\mathcal{P}_{\rm NNLO}}{\mathcal{P}_0}
&=& \frac{7}{4}\frac{d_F}{d_A}\frac{1}{N_f}\lb N_f+\frac{240}{7}\hmu_I^2
+\frac{480}{7}\hmu_I^4\rb
    -\frac{s_F\alpha_s}{\pi}\frac{1}{N_f}\bigg[\frac{5}{8}
\left(5N_f+144\hat\mu_I^2 + 288\hat\mu_I^4\right)
    \nn
    &-&\frac{15}{2}\left(N_f + 24\hat\mu_I^2\right)\hat m_D - 
\frac{15}{2}\bigg(2N_f\ln{\frac{\hat\Lambda}{2}-N_f
  +2(2\log2+\gamma_E) -2\aleph(z_I)}\Big)\hat m_D^3
      +90\hat m_q^2 \hat m_D\bigg]
\nn
&+& \frac{s_{2F}}{N_f}\left(\frac{\alpha_s}{\pi}\right)^2\bigg[\frac{15}{64}
\bigg\{35N_f - 32\log2 - 
     32\lb N_f-1-24\hmu_I^2\rb\frac{\zeta'(-1)} {\zeta(-1)} + 
944 \hat\mu_I^2+2656 \hat\mu_I^4\nn
      &-& 384\Big(12i\hat\mu_I\aleph(2,z_I) - 2(1+8\hat\mu_I^2)\aleph(1,z_I) 
- i\hat\mu_I(1+4\hat\mu_I^2)
      \aleph(0,z_I)\Big)\bigg\} - \frac{45}{2}\hat m_D
\left(N_f+24\hat\mu_I^2\right)\bigg] \nonumber
\eqa
\bqa
&+& \left(\frac{s_F\alpha_s}{\pi}\right)^2
      \frac{1}{N_f}\frac{5}{16}\Bigg[96\left(N_f+24\hat\mu_I^2\right)
\frac{\hat m_q^2}{\hat m_D}
     +\frac{4}{3}\lb5N_f+144\hat\mu_I^2+288\hat\mu_I^4\rb
      \ln\frac{\hat{\Lambda}}{2}+\frac{N_f}{3}\(1+12\gamma_E\)\nn
    &+& 16(7+12\gamma_E)\hat\mu_I^2+224\hmu_I^4-
\frac{8}{15N_f}\(8N_f^2-21N_f+54\)\frac{\zeta^{\prime}(-3)}{\zeta(-3)}-
   \frac{16}{N_f}\(N_f(1+16\hat\mu_I^2)-12\hmu_I^2\)
\frac{\zeta^{\prime}(-1)}{\zeta(-1)}\nn
   &-&\frac{8}{5N_f}(11N_f+120\hmu_I^2)\log2
   -  192\Big\{8\aleph(3,z_I) + 12i\hat\mu_I\aleph(2,z_I)-2(1+2\hat\mu_I^2)
\aleph(1,z_I)-i\hat\mu_I\aleph(0,z_I)\Big\}\Bigg] \nn
&+& \left(\frac{s_F\alpha_s}{\pi}\right)^2
      \frac{1}{N_f^2}\Bigg[\frac{5}{4\hat m_D}\left(N_f+24\hat\mu_I^2\right)^2
     +360\Bigg\{ 2\left(1 +\gamma_E\right)\hat\mu_I^4 - \aleph(3,2z_I)-2
\aleph(3,z_I+z_0) \nn
     &-& 4i\hat\mu_I\left[\aleph(2,2z_I)+\aleph(2,z_I+z_0)\right]
      +4\hmu_I^2\aleph(1,z_I)
       +2\hat\mu_I^2(2\aleph(1,2z_I) + \aleph(1,z_I+z_0)) + 4i\hat\mu_I^3
\aleph(0,z_I)\Bigg\}\nn
       &-&\frac{15}{2}\lb N_f+24\hat\mu_I^2\rb\lb 2N_f\L-N_f +2(2\log2+\gamma_E)
-2\aleph(z_I)\rb  \hat m_D\Bigg]\nonumber\\
       &+& \left(\frac{c_A\alpha_s}{3\pi}\right)
\left(\frac{s_F\alpha_s}{\pi N_f}\right)
\Bigg[\frac{15}{2\hat m_D}\lb N_f + 24\hmu_I^2\rb
     -\frac{235}{16}\Bigg\{\bigg(N_f + \frac{1584}{47}\hat\mu_I^2+
\frac{3168}{47}\hat\mu_I^4\bigg)\ln\frac{\hat\Lambda}{2}
     \nonumber\\
    &-&\frac{144}{47}\lb N_f+24\hmu_I^2\rb\log\hat m_D+\frac{319}{940}
\left(N_f+\frac{4080}{319}\hat\mu_I^2+\frac{77280}{319}\hat\mu_I^4\right)
   -\frac{24\gamma_E }{47}\lb N_f+24\hat\mu_I^2\rb
\nonumber\\
    &-&
   \frac{2}{47}\lb 22N_f+15+624\hmu_I^2\rb\frac{\zeta'(-1)}{\zeta(-1)}
    -\frac{1}{235}(268N_f-273)\frac{\zeta'(-3)}{\zeta(-3)} + 
\frac{111}{235}\log2\nn
  & -&\frac{144}{47}\Big[4i\hat\mu_I\aleph(0,z_I)+\left(5-92\hat\mu_I^2\right)
\aleph(1,z_I) + 144i\hmu_I\aleph(2,z_I)
   + 52\aleph(3,z_I)\Big]\Bigg\}\nn
   &+& 90\frac{\hat m_q^2}{\hat m_D}+\frac{315}{4}\Bigg\{\lb N_f+\frac{264}{7}
\hmu_I^2\rb\L + \frac{11}{7}\lb N_f+24\hmu_I^2- \frac{4}{11}\rb\gamma_E
\nonumber\\
   &+& \frac{9}{14}\lb N_f+\frac{264}{9}\hmu_I^2\rb
+\frac{2}{7}\left(2\aleph(z_I) - 4\ln2\right)\Bigg\}\hat m_D 
\Bigg]
+ \frac{\Omega_{\rm NNLO}^{\rm YM}}{\Omega_0} \, ,
\label{finalomega2}
\end{eqnarray}

In Fig.~\ref{pres1}, we show the NNLO pressure obtained using
HTLpt as a function of
$T$ normalized to that of an ideal gas of massless particles
for $\mu_I=200$ MeV, $\mu_B=0$  (left) and $\mu_I=200$ MeV, $\mu_B=400$ Mev 
(right). 
The pressure is an increasing function of $T$, but stays well below
the ideal-gas value even for the highest temperatures shown.


\begin{figure}[htb]
\subfigure{\hspace{-2mm}\includegraphics[width=7.5cm]{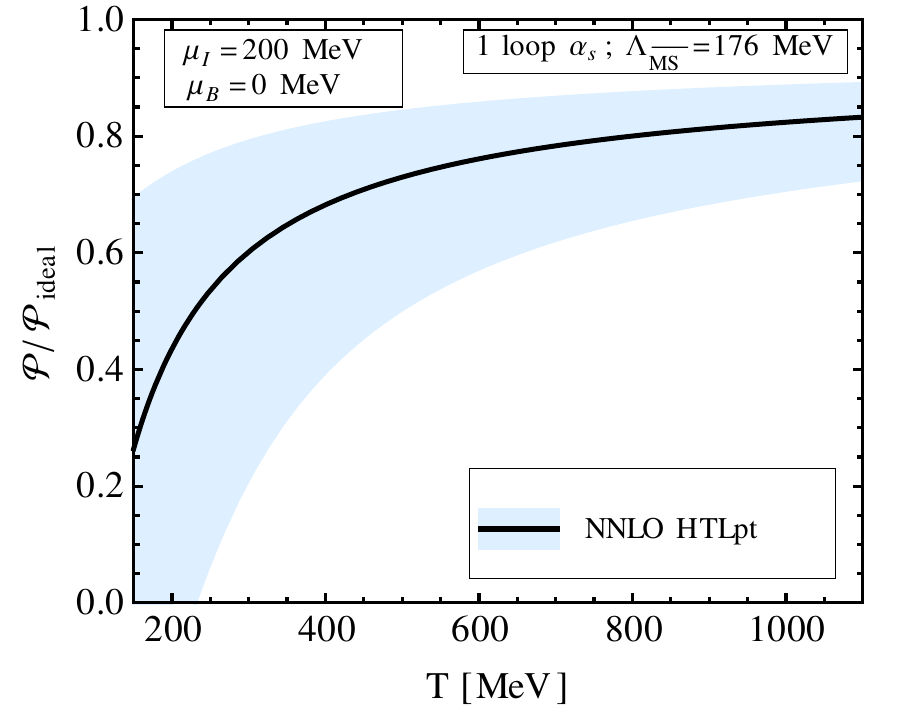}} 
\subfigure{\includegraphics[width=7.5cm]{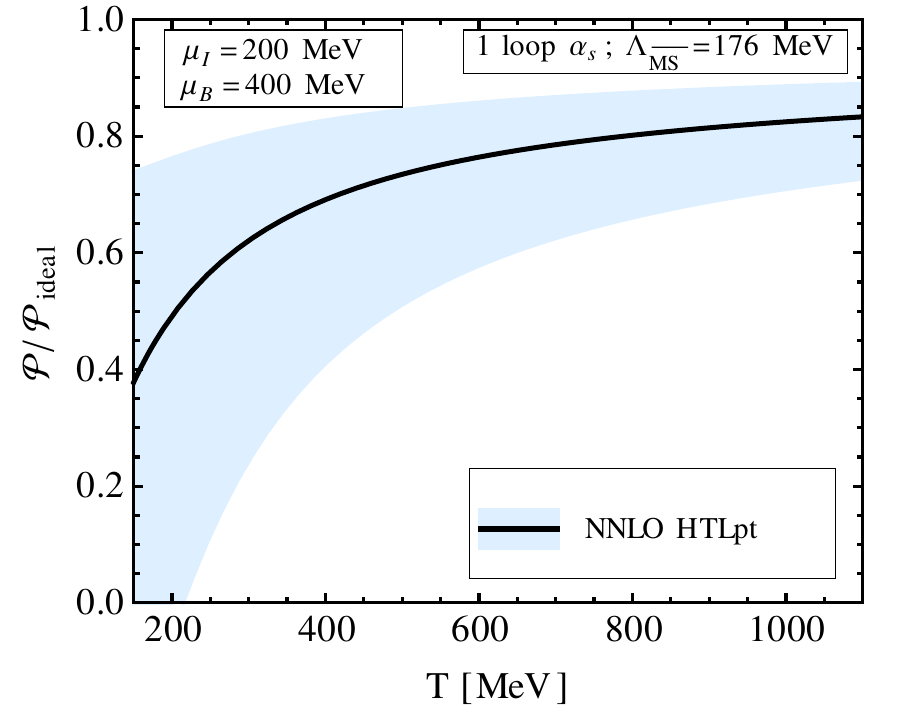}}
\caption{The pressure normalized to that of an ideal gas of massless particles
as a function of $T$. Left figure is for $\mu_I=200$ Mev, $\mu_B=0$ and right figure is 
for $\mu_I=200$ MeV, $\mu_B=400$ MeV.}
\label{pres1}
\end{figure}

In Fig.~\ref{pres2}, we show the normalized 
NNLO pressure of HTLpt as a function of $T$ for four different values
of the isospin chemical potential $\mu_I$.
We notice that the pressure is an increasing function of $\mu_I$
for fixed temperature and that the pressure curves converge at a temperature
of approximately 800 MeV.

\begin{figure}[htb]
\begin{center}
\subfigure{\hspace{-2mm}\includegraphics[width=7.5cm]{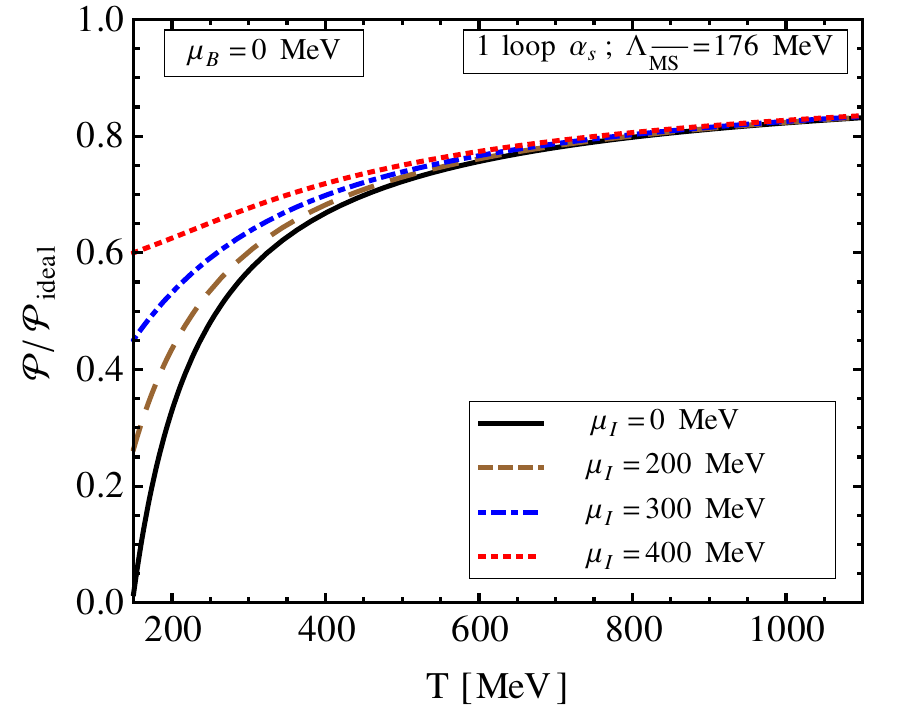}}
\subfigure{\includegraphics[width=7.5cm]{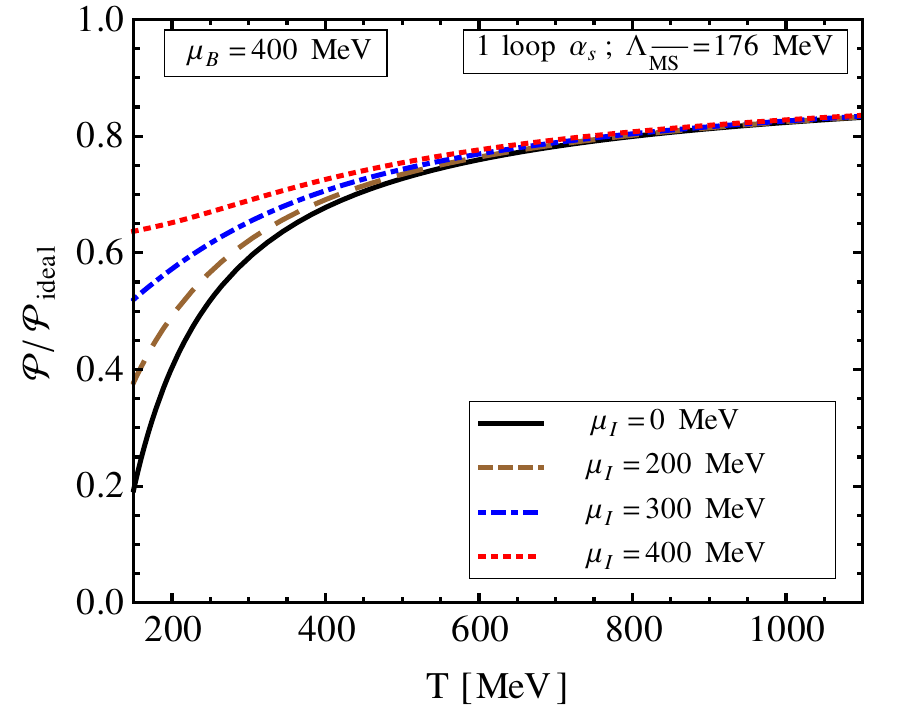}}
\caption{The pressure normalized to that of an ideal gas of massless particles
as a function of $T$ for various values of the isospin chemical potential
$\mu_I$ at $\mu_B=\mu_S=0$ (left) and $\mu_B=400$ MeV, $\mu_S=0$ (right). 
Here 
$\Lambda_g=2\pi T$ and $\Lambda_q= 2\pi \sqrt{T^2+(\mu_B^2+2\mu_I^2)/(3\pi^2)}$ 
were used.
}
\label{pres2}
\end{center}
\end{figure}

\end{widetext}

\subsection{Energy density}
\begin{widetext}

\begin{figure}[htb]
\subfigure{\hspace{-2mm}\includegraphics[width=7.5cm]{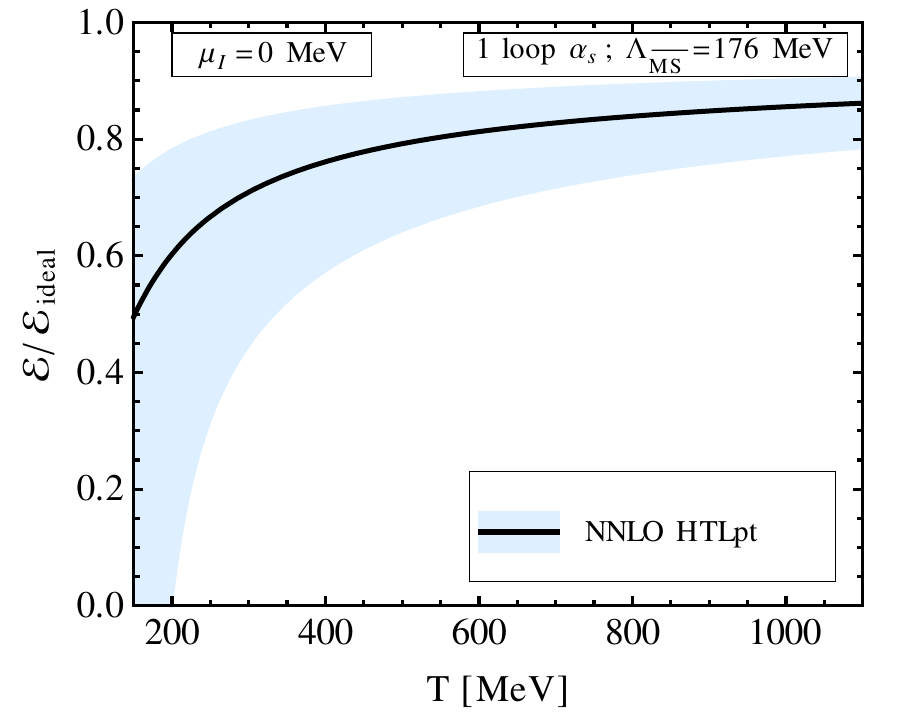}} 
\subfigure{\includegraphics[width=7.5cm]{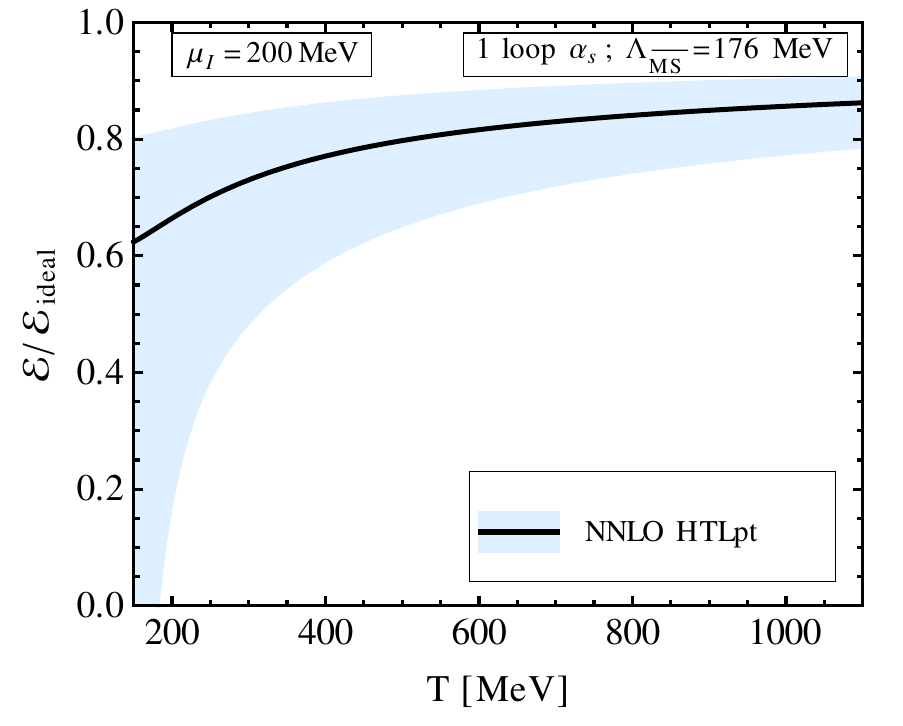}} 
\caption{The energy density 
normalized to that of an ideal gas of massless particles
as a function of $T$. Left figure is for $\mu_I=0$ and right figure is 
for $\mu_I=200$ MeV. $\mu_B=\mu_S=0$ in both plots. 
}
\label{energy1}
\end{figure}

\end{widetext}

Once we know the pressure ${\cal P}$, we can calculate the energy density
${\cal E}$ by the Legendre transform
\bqa \nonumber
{\cal E}&=&
T{\partial{\cal P}\over\partial T}
+\mu_q{\partial{\cal P}\over\partial\mu_q}-{\cal P}
\\ 
&=&T{\partial{\cal P}\over\partial T}
+\mu_I{\partial{\cal P}\over\partial\mu_I}-{\cal P}\;.
\eqa
where we have used that $\mu_I={1\over2}(\mu_u-\mu_d)$ and
$\mu_s=0$.
In Fig.~\ref{energy1}, we show the energy density as a function of 
the temperature for $\mu_I=0$ (left) and $\mu_I=200$ MeV (right).
As in the case of the pressure, the energy density is an increasing
function of $T$ and stays well below the ideal-gas value
for all temperatures.

\begin{figure}[htb!]
\begin{center}
\includegraphics[width=7.5cm]{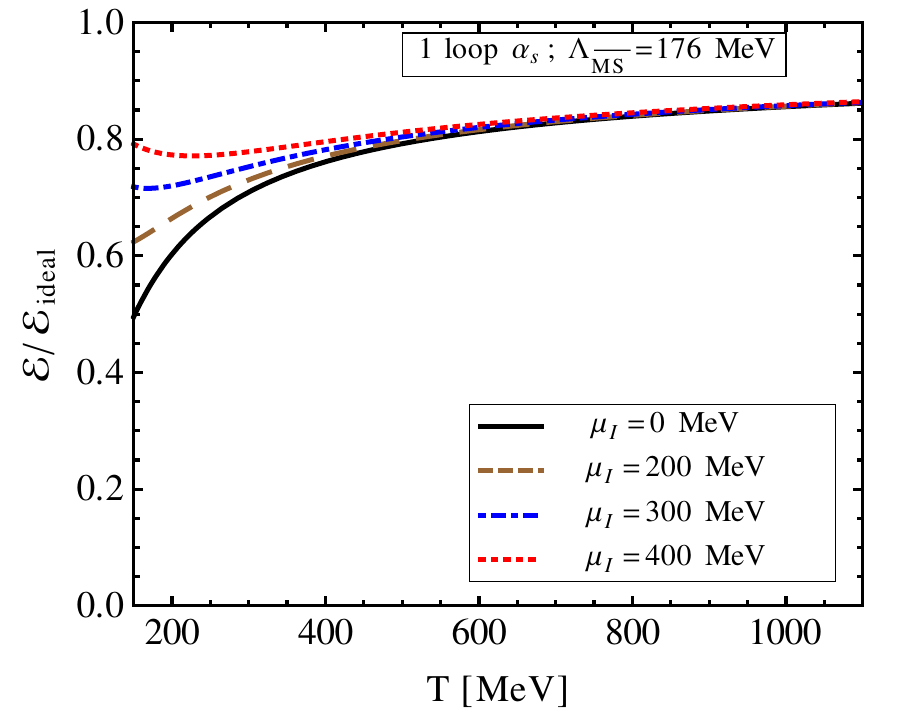} 
\end{center}
\caption{The energy density 
normalized to that of an ideal gas of massless particles
as a function of $T$ for four different values of the isospin chemical 
potential and $\mu_B=\mu_S=0$. Here 
$\Lambda_g=2\pi T$ and $\Lambda_q= 2\pi \sqrt{T^2+2\mu_I^2/(3\pi^2)}$
were used.}
\label{energy2}
\end{figure}

In Fig.~\ref{energy2}, we show the normalized
energy density for four different values a
of the isospin chemical potential $\mu_I$. 
For $\mu_I=0$ or $\mu_I=200$, the energy density is 
an increasing function of $T$. Note, however, that there is
a minimum for the energy density for low temperatures and
higher values of the isospin
chemical potential. We would like
to mention here that HTLpt probably cannot be trusted at these low  
temperatures with large chemical potential and one can not 
attribute any interesting physics to this nonmonotonic behavior.

Likewise, the curves converge at high temperatures,
here already at approximately $T=600$ MeV. 

\subsection{Trace anomaly}
\begin{figure}
\subfigure{\hspace{-2mm}\includegraphics[width=7.5cm]{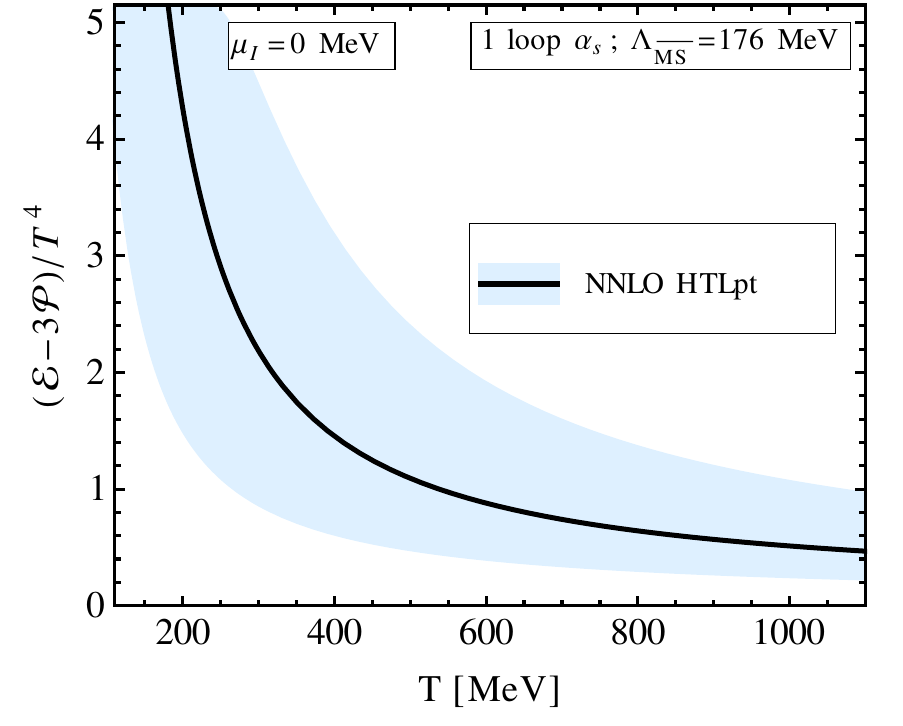}} 
\subfigure{\hspace{-2mm}\includegraphics[width=7.5cm]{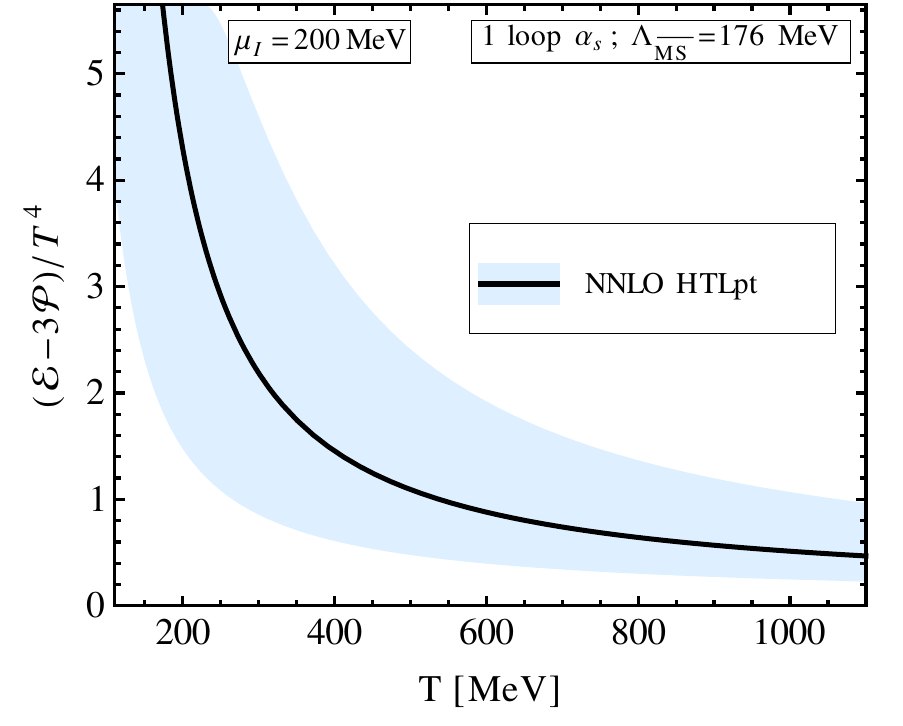}} 
\caption{Trace anomaly divided by $T^4$ as a function of the temperature $T$.
Left figure is for $\mu_I=0$ and right figure is for $\mu_I=200$ MeV.
$\mu_B=\mu_S=0$ in both plots. 
}
\label{anom1}
\end{figure}

The trace anomaly or interaction measure ${\cal I}$
is defined by the difference
\bqa
{\cal I}&=&{\cal E}-3{\cal P}\;.
\eqa
For an ideal gas of massless particles, the trace anomaly vanishes
since ${\cal E}=3{\cal P}$. For massless particles and nonzero $g$,
${\cal I}$ is nonzero and is a measure of the interactions in the
plasma.\footnote{For nonzero current quark masses $m_0$, ${\cal I}\neq0$
even in the absence of interactions.}
In Fig.~\ref{anom1}, we show the interaction measure as a function of
the temperature for two different values of the isospin chemical potential,
$\mu_I=0$ (left) and $\mu_I=200$ MeV (right).
The trace anomaly is a decreasing function of $T$ and it converges
to zero for large values of $T$ due to asymptotic freedom.

\begin{figure}
\begin{center}
\includegraphics[width=7.5cm]{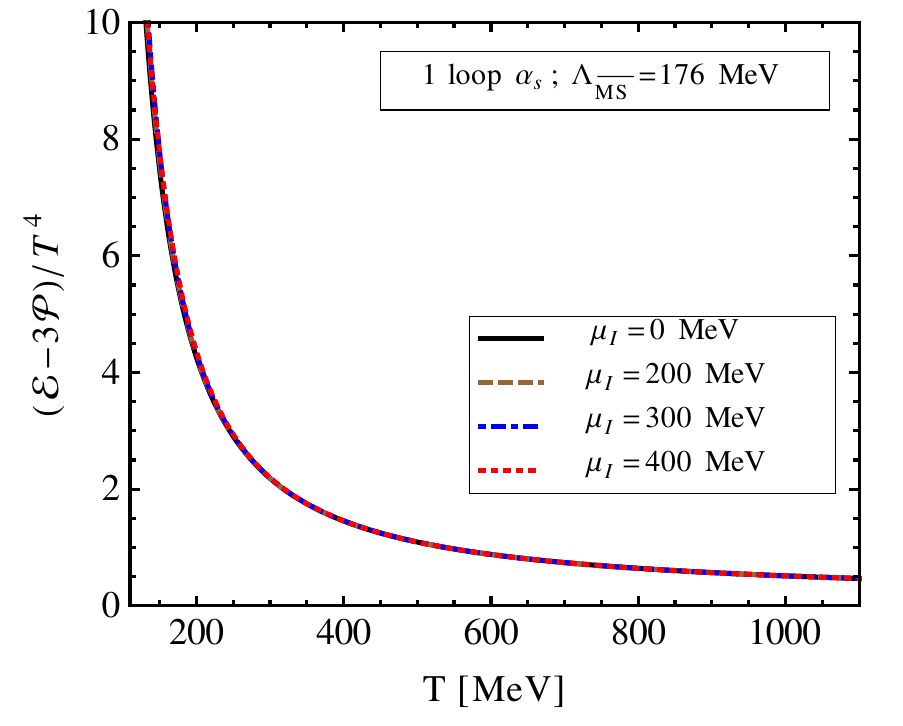} 
\end{center}
\caption{Trace anomaly divided by $T^4$ as a function of the temperature $T$
for four different values of the isospin chemical potential
and $\mu_B=\mu_S=0$. 
Here $\Lambda_g=2\pi T$ and $\Lambda_q= 2\pi \sqrt{T^2+2\mu_I^2/(3\pi^2)}$
were used.}
\label{anom2}
\end{figure}

In Fig.~\ref{anom2}, we show the normalized interaction measure 
as a function of the temperature $T$ for four different values of the
isospin chemical potential $\mu_I$. As the figure demonstrates, 
the curves are essentially identical.

\subsection{Speed of sound}
\begin{widetext}
The speed of sound $c_s$ is defined by
\bqa
c_s^2&=&{\partial{\cal P}\over\partial{\cal E}}\;.
\eqa
In Fig.~\ref{sound1}, we show the speed of sound squared $c_s^2$ for
two different values of the isospin chemical potential, $\mu_I=0$ (left)
and $\mu_I=200$ MeV (right). 
The horizontal dotted line is the
ideal-gas value $c_s^2={1\over3}$. 
As this figure demonstrates, the speed of sound is an increasing
function of $T$.

\begin{figure}[htb!]
\hspace{-2mm}\includegraphics[width=7.5cm]{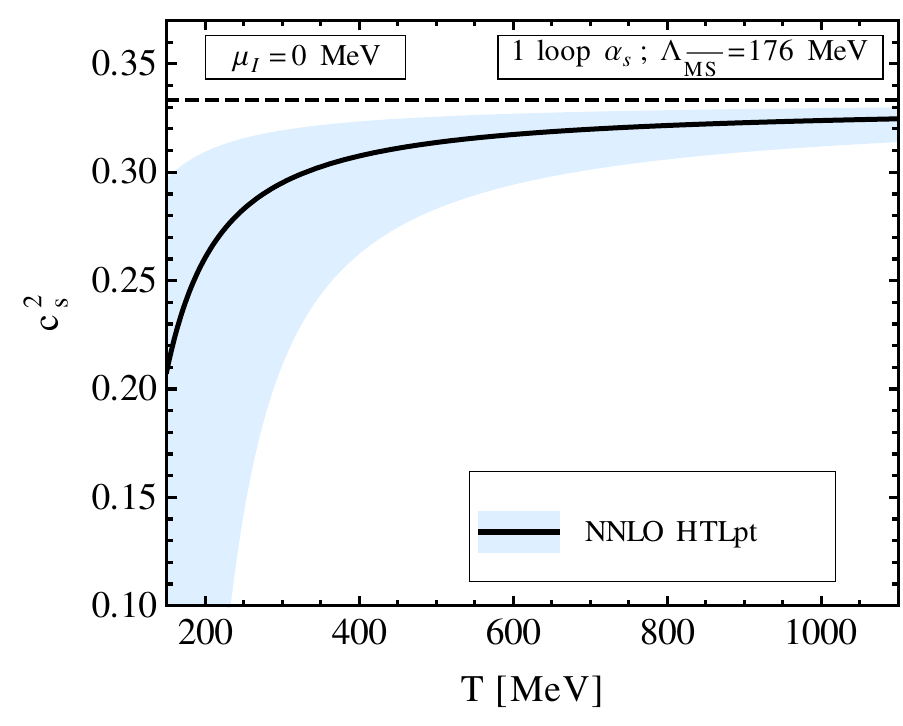}
\hspace{1cm}
\includegraphics[width=7.5cm]{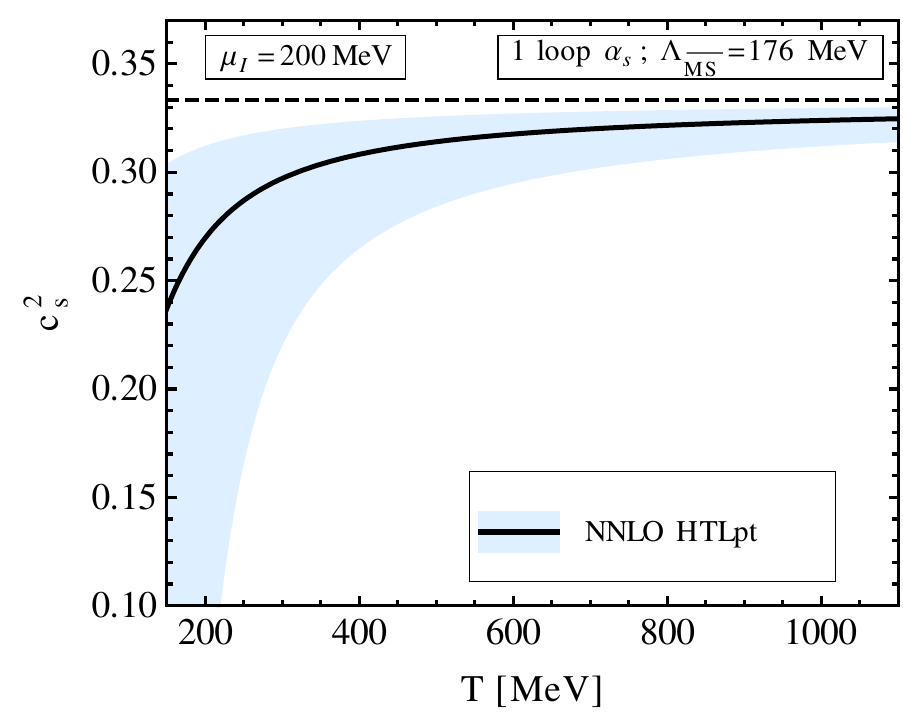}
\caption{Speed of sound squared as a function of the temperature $T$.
Left figure is for $\mu_I=0$ and right figure is for $\mu_I=200$ MeV.
$\mu_B=\mu_S=0$ in both plots.}
\label{sound1}
\end{figure}

\end{widetext}

\begin{figure}[htb]
\begin{center}
\includegraphics[width=7.5cm]{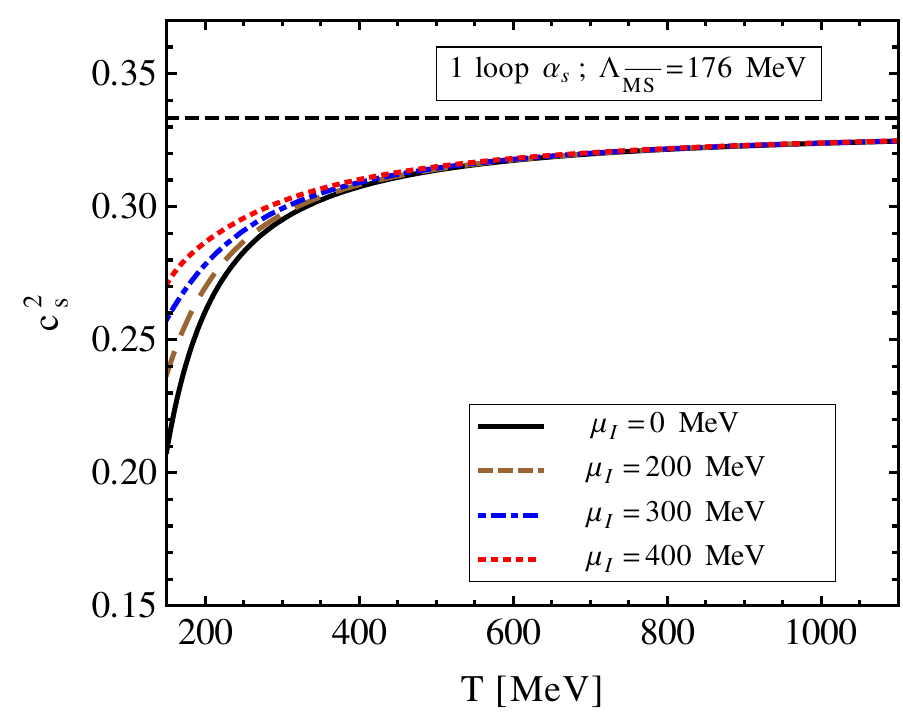}
\end{center}
\caption{Speed of sound squared as a function of the temperature $T$
for different values of the isospin chemical potential $\mu_I$ and
$\mu_B=\mu_S=0$.
Here 
$\Lambda_g=2\pi T$ and $\Lambda_q= 2\pi \sqrt{T^2+2\mu_I^2/(3\pi^2)}$
were used.}
\label{sound2}
\end{figure}

In Fig.~\ref{sound2}, we show the speed of sound squared 
$c_s^2$ for four different
values of the isospin chemical potential $\mu_I$. We notice that the
speed of sound is an increasing function of $\mu_I$ for fixed $T$
and that the curves converge rather quickly, here at approximately 
$T=400$ MeV.


\subsection{Susceptibilities}
Using the thermodynamic potential given by 
Eq.~(\ref{finalomega2}), we can compute the quark-number susceptibilities.
In the most general case, we have one quark chemical potential $\mu_f$
for each quark flavor $f$, which we can organize in an
$N_f$-dimensional vector $\boldsymbol\mu=(\mu_u,\mu_d,\mu_s,...,\mu_{N_f})$.
The single quark susceptibilities are defined by
\bqa
\chi_{ijk...}(T)&=&
{\partial^{i+j+k+...}{\cal P}(T,{\boldsymbol\mu})\over\partial\mu_u^i
\partial\mu_d^j\partial\mu_s^k...}
\bigg|_{{\boldsymbol \mu}={\boldsymbol \mu_0}}\;,
\eqa
where $\boldsymbol\mu_0$ is a configuration of quark chemical potentials.
When computing the derivatives with respect to the chemical
potential, we 
will use $\boldsymbol\mu_0=\boldsymbol 0$.
We treat $\Lambda_q$ as being a constant and only put the chemical 
potential dependence of $\Lambda_q$ in after the
derivatives are taken. We have done this in order to more closely 
match the procedure
used to compute the susceptibilities using resummed dimensional 
reduction~\cite{vuorinen1}
and to ensure that the susceptibilities vanish when $N_f=0$.
In the following, we will use a shorthand notation for the quark
susceptibilities by specifying derivatives by a string of quark flavors
using superscript. For example, $\chi^{uu}_{2}=\chi_{200}$, 
$\chi^{ds}_{2}=\chi_{011}$, and $\chi^{uudd}_{4}=\chi_{220}$.
For a three-flavor system with $(u,d,s)$ quarks
with $\mu_B=\mu_S=0$
, the 
n'th-order isospin number susceptibility 
evaluated at $\mu_I=0$ is defined by
\bqa
\chi^I_n&\equiv&{\partial^n{\cal P}\over\partial\mu_I^n}\bigg|_{\substack{\hmu_I=0}}
\;.
\eqa
We can analytically express various order susceptibilities as
\begin{widetext}
\bqa
\chi^I_2
&=&\frac{1}{48}T^2\left(32N_c-3d_A\hmD^2\hmD''\right) 
- \frac{d_A\alpha_s T^2}{96\pi}\Bigg[16(3 - 6\hmD + 7\hmD^3\zeta(3)) - 
\Bigg\{2(2c_A + N_f) - 24c_A\hmD\nn
&-&\left(15 c_A - 36\gamma_E + 66 c_A\gamma_E + 6 N_f - 72\log2 - 36 
\L + 66 c_A \L\right) \hmD^2\Bigg\}\hmD''\Bigg] \nn
&+&\frac{d_A\alpha_s^2T^2}{144\pi^2}\Bigg[\frac{24(3+2c_A)}{\hmD}+360 - 91c_A + 
105c_F +12c_A\gamma_E-146N_f+24\gamma_E N_f\nn
&+& (- 504 + 144 c_A + 96 c_F + 168 N_f)\Za -(72 + 112 c_A - 48 c_F + 8 N_f)
\log2\nonumber
\eqa
\bqa
&-& (132c_A - 24 N_f)\log\frac{\hat\Lambda}{2} + 288c_A\log\hmD - 
12\Big(36c_F + 36\gamma_E - 6 N_f + 72\log2\nn
&-& 6(11 c_A - 2N_f)\log\frac{\hat\Lambda}{2} - 21\zeta(3) - c_A(33 + 
66\gamma_E + 14\zeta(3))\Big)\hmD\nn
&-&\Bigg\{\frac{(3+2c_A)^2}{4\hmD^2} + \frac{6(3+2c_A)}{\hmD} + \frac{3}{4}
\Big(3 c_A (9 + 14\gamma_E - 16\log2) 
- c_A^2 (79 - 44\gamma_E + 4 \pi^2 - 44\log2) \nn
&-&  6 (6 c_F + 6\gamma_E - N_f + 12\log2) + 
 2 (21 c_A + 22 c_A^2 - 6 N_f)\L\Big)\Bigg\}\hmD^{\prime} 
\Bigg]\;,
\eqa
\bqa
\chi^I_4&=&\frac{1}{\pi^2}\left(4N_c-\frac{d_A\hmD}{64}
\left(6\hmD''^2+\hmD\hmD^{iv}\right)\right) 
- \frac{d_A\alpha_s}{384\pi^3}\Bigg[1152 - 5952 \hmD^3 \zeta(5) -
   \Big(576 - 2016 \hmD^2 \zeta(3)\Big) \hmD''\nn 
&& +\left\{72 c_A -\left(  216 \gamma_E - 90 c_A - 396 c_A \gamma_E  - 
36 N_f + 432 \log2 + 216 \L -396 c_A\L\right)\hmD\right\}\hmD''^2  \nn
&& - \left\{4 c_A + 2 N_f - 24 c_A \hmD - \left(15 c_A + 6 N_f - 
    72 \log2 - 6(6-11 c_A)\left(\gamma_E+ \L\right)\right)\hmD^2\right\} 
\hmD^{iv}\Bigg]\nn
&&+\frac{d_A\alpha_s^2}{24\pi^4}\Bigg[\frac{144}{\hmD} -36 + 21c_F(5+4\zeta(3)) 
+ 72 \gamma_E +
    2 N_f\(11+14\zeta(3)+12\L\) - 576 \log2   \nn
&& - 11 c_A \(17 + 12\gamma_E + 24\log2 - 7 \zeta(3)+12\L\) + 
 6 \Big(168 \zeta(3) - 93 \zeta(5) - 62 c_A \zeta(5)\Big)\hmD\nn
&&+\Bigg\{-\frac{6 (3 + 2 cA)}{\hmD^2} + \frac{72 c_A}{\hmD} + 
 3 (-36 c_F - 36 \gamma_E + 6 N_f - 72 \log2 + 6 (11 c_A - 2 N_f)\L\nn
 &&   + 21 \zeta(3) + c_A (33 + 66 \gamma_E + 14 \zeta(3))\Bigg\}\hmD'' + 
\frac{(3 + 2 c_A)}{16\hmD^3}\Bigg\{(3 + 2 c_A) - 12c_A\hmD\Bigg\}\hmD''^2\nn
&&-\Bigg\{\frac{(3 + 2 c_A)^2}{96 \hmD^2} - \frac{c_A(3 + 2 c_A)}{4 \hmD} - 
 \frac{1}{32}\Bigg(3 c_A (9 + 14 \gamma_E - 16 \log2) + 
    c_A^2 (-79 + 44 \gamma_E - 4 \pi^2 + 44 \log2) \nn
    && - 6 (6 c_F + 6 \gamma_E - N_f + 12\log2) + 
    2 (21c_A + 22 c_A^2 - 6 N_f)\L\Bigg)\Bigg\}\hmD^{iv}
 \Bigg]\;,
\eqa
where
\bqa
\hmD''&=& \left.\frac{\partial^2\hmD}{\partial\hmu_I^2}\right|_{\substack{\hmu_I=0
\\ \hmu_B=0}}\nn
&=&\frac{\alpha_s}{9\pi^2\hmD} \Bigg[ 36 \pi + \alpha_s \Bigg\{6(11N_c - 2N_f) 
\L - 54 c_F + 
      N_f (6 - 12\gamma_E - 24\log2 + 7 \zeta(3)) \nn
      && \hspace{2cm} + N_c (33 + 66 \gamma_E + 14 \zeta(3))\Bigg\}\Bigg]\;,
\eqa
\bqa
\hmD^{iv}&=& \left.\frac{\partial^4\hmD}{\partial\hmu_I^4}
\right|_{\substack{\hmu_I=0\\ \hmu_B=0}}\nn
&=&-\frac{\alpha_s^2}{54\pi^4\hmD^3} \Bigg[ 2\Bigg\{36 \pi + \alpha_s 
\Big( 6(11 c_A - 2 N_f)\L -54 c_F + 
        N_f(6 - 12 \gamma_E - 24 \log2 + 7 \zeta(3)) \nn
        && + c_A(33 + 66 \gamma_E + 14 \zeta(3))\Big)\Bigg\}^2
        - \alpha_s \Bigg\{12\pi(2c_A + N_f) + \alpha_s\Big(2(5 + 22 \gamma_E) 
c_A^2  + 
     3 c_A (9 + 14 \gamma_E - 16 \log2)\nn
     && - 6 (9 c_F + N_f  (-1 + 2 \gamma_E + 4\log2)) + 
     2(21 c_A + 22 c_A^2 - 6 N_f)\L\Big)\Bigg\} \big(84 N_f \zeta(3) - 
31(2c_A+N_f) \zeta(5)\big)\Bigg]\;.\nn 
\eqa
\end{widetext}

For a three-flavor system consisting of $(u,d,s)$ quarks, we can express the
isospin susceptibilities in terms of the quark susceptibilities as

\bqa
\label{susp1}
\chi^I_2&=&
\left[\chi^{uu}_2+\chi^{dd}_2-2\chi^{ud}_2\right]
\;,
\\
\chi^I_4&=&
\left[\chi^{uuuu}_4+\chi^{dddd}_4-4\chi^{uuud}_4
-4\chi^{dddu}_4
+6\chi^{uudd}_4
\right]\;.
\label{susp3}
\eqa

\begin{figure}[htb]
\centering
\subfigure{\hspace{-2mm}\includegraphics[width=7.5cm]{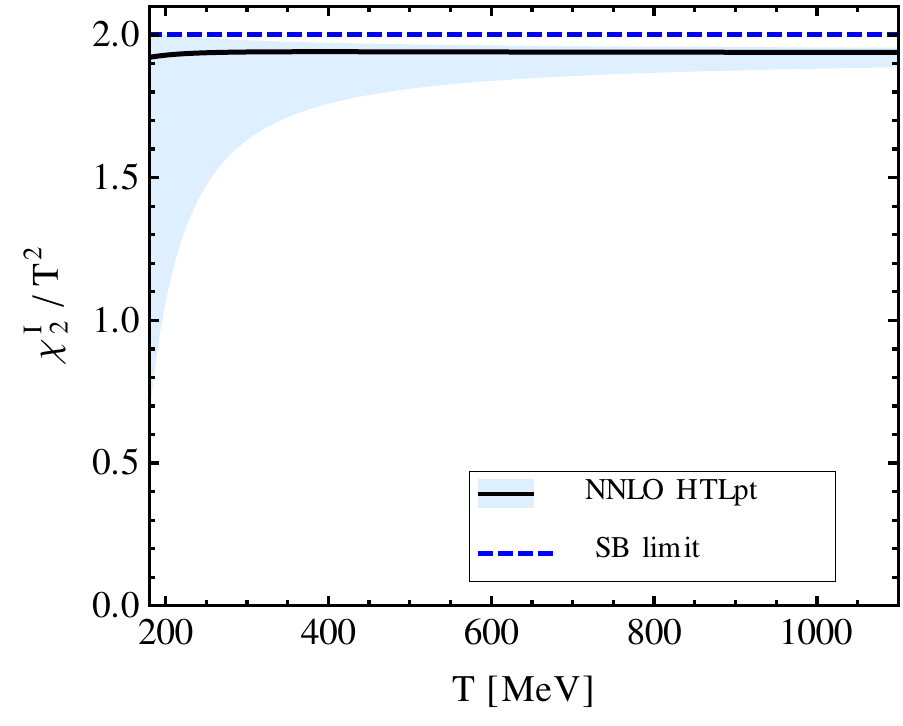}} 
\hspace{1cm}
\subfigure{\includegraphics[width=7.5cm]{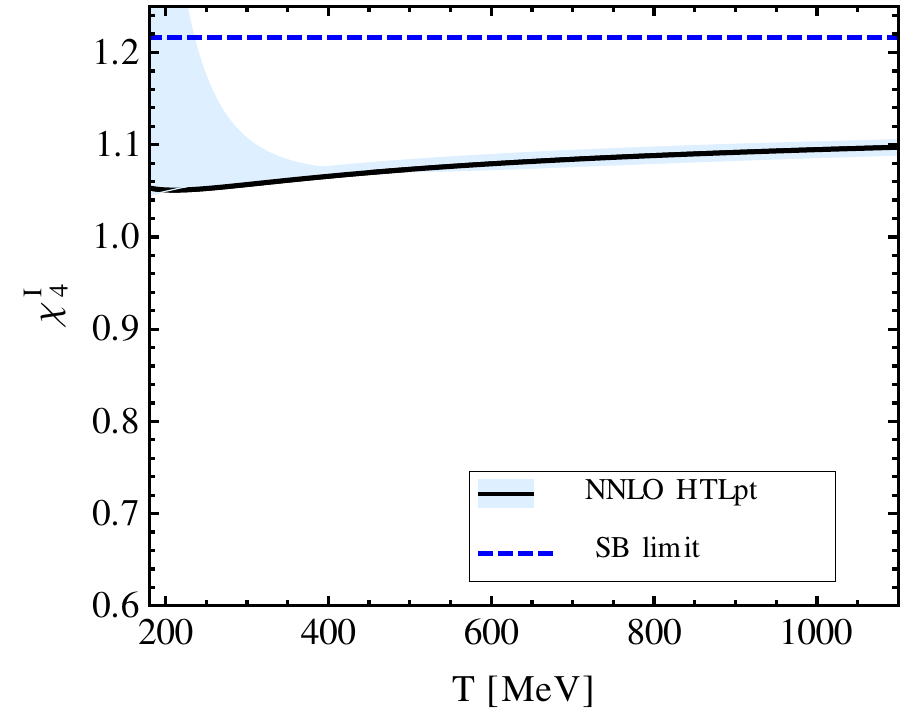}}
\caption{Second and fourth-order
susceptibilities as functions of the
temperature $T$ normalized to 
$T^2$ and one, respectively.}
$\mu_I=\mu_B=\mu_S=0$ in both plots. 
\label{susc1}
\end{figure}

The isospin susceptibilities are expressed in terms of
diagonal (same flavor on all indices) quark susceptibilities
or off-diagonal  (different flavor on some or all indices).
In HTLpt, there are off-diagonal susceptibilities arising explicitly
from some of the three-loop graphs~\cite{3loopqcd2,complete}. 
There also potential off-diagonal contributions coming from all HTL
terms since the mass parameter $m_D$ receives contributions from 
all quark flavors. However, 
these contributions vanish
when we evaluate the susceptibilities at $\mu_f=0$.
In this case, the HTLpt second and fourth-order
isospin susceptibilities reduce to
\bqa
\chi^I_2&=&2\chi^{uu}_2\;,\\
\chi^I_4&=&\left[
2\chi^{uuuu}_4+6\chi^{uudd}_4\right]\;.
\eqa
In Fig.~\ref{susc1}, we show the HTLpt predictions for the isospin
second and  fourth-order
susceptibilities $\chi_2^I/T^2$ and  $\chi_4^I$
as functions of $T$.
The horizontal dotted lines are the corresponding 
isospin susceptibilities for an ideal gas, indicated by Stefan-Boltzmann
limit.
The central line for the
second-order susceptibility is almost flat, while the 
central line for the and fourth-order susceptibility
is slowly increasing.

%

\section{Summary}
\label{conclude}
In this paper, we presented results for a number of thermodynamic functions of 
QCD at finite temperature $T$ 
and finite isospin chemical potential $\mu_I$ using 
hard-thermal-loop perturbation theory. 
The pressure was also calculated at nonzero baryon chemical potential
$\mu_B$. Our results were derived from the 
three-loop thermodynamic potential, which was computed in Ref.~\cite{complete} 
as a 
function of temperature and quark chemical potentials.  Our final results 
depend on two renormalization scales $\Lambda_g$ and $\Lambda_q$ which are 
expected to be approximately $2 \pi T$ and 
$2\pi\sqrt{T^2 + (\mu_B^2+2\mu^2_I)/(3\pi^2)}$.
In order to gauge the theoretical uncertainty associated with the scale choice, 
we varied both $\Lambda_g$ and $\Lambda_q$ by a factor of two (light-blue bands 
in some figures).  We found that most quantities have a sizable scale variation 
and, at this moment in time, we do not have a method to reduce the size of the 
bands. 
A solution to this problem is suggested by the 
authors of Ref.~\cite{kneur}. 
In this approach, dubbed renormalization group optimized perturbation theory,
the authors modify standard optimized perturbation theory or
SPT. This is done by changing the added/subtraction mass term, including
a finite vacuum term, and imposing renormalization group invariance
on the pressure.
In the case of $\phi^4$-theory, the result for the pressure
up to two-loop order is very stable and with narrow bands
under a scale variation.
Note, however, that some 
quantities, e.g. $\chi_4^I$, have very small scale variation for temperatures 
$T\gtrsim$ 400 MeV and hence HTLpt provides testable predictions.

Given the relatively good agreement between lattice results and the predictions 
of NNLO HTLpt at zero and finite baryon chemical potential for $T \gtrsim 250$ 
MeV, we expect that the lattice results at finite $\mu_I$ should fall close to 
the central (black) lines predicted herein at high temperatures.  We are looking
forward to lattice measurements of QCD thermodynamics at finite $\mu_I$ and high
temperatures (with $\mu_B=\mu_S=0$)
in order to test the predictions made herein.  
Since the necessary 
lattice measurements can be done without Taylor expansion, they would provide a 
high-precision test of NNLO HTLpt.

\section*{Acknowledgments}
N.~Haque was supported by an award from the Kent State University Office of 
Research and Sponsored Programs.  
M.~G. Mustafa was supported by the Indian Department of Atomic Energy under the 
project "Theoretical Physics across the energy scales".
M.~Strickland was supported by the U.S. Department
of Energy under Award No.~DE-SC0013470.

\bibliography{refs}{}

\begin{thebibliography}{99}




\bibitem{kajantie} 
K. Kajantie, M. Laine,  K. Rummukainen and Y. Schroder, 
Phys. Rev. D {\bf 67}, 105008 (2003).

\bibitem{vuorinen1}
A.~Vuorinen, 
Phys.\ Rev.\ D {\bf 67},  074032 (2003).

\bibitem{vuorinen2}
A. Vuorinen,
Phys.\ Rev.\ D {\bf 68}, 054017  (2003).



\bibitem{ipp} A. Ipp, K. Kajantie, A. Rebhan, and A. Vuorinen, 
\emph{Phys. Rev. D}{\bf 74}, 045016 (2006).









\bibitem{spt1}
F. Karsch, A. Patkos and P. Petreczky, 
Phys. Lett. B {\bf 401}, 69 (1997).

\bibitem{spt2}
S. Chiku and T. Hatsuda, 
Phys. Rev. D {\bf 58}, 076001 (1998).

\bibitem{spt3}
J.O. Andersen, E. Braaten and M. Strickland, 
Phys. Rev. D {\bf 63}, 105008 (2001).


\bibitem{spt4}
J.O. Andersen and L. Kyllingstad, 
Phys. Rev. D {\bf 78}, 076008 (2008).


\bibitem{vpt1}
V.I. Yukalov, 
Teor. Mat. Fiz. {\bf 26} 403,
(1976).

\bibitem{vpt2}
P.M. Stevenson, 
Phys. Rev. D {\bf 23}, 2916 (1981).

\bibitem{vpt3}
A. Duncan and M. Moshe, 
Phys. Lett. B {\bf 215}, 352 (1988).

\bibitem{vpt4}
A. Duncan and H.F. Jones, 
Phys. Rev. D {\bf 47}, 2560 (1993).

\bibitem{vpt5}
A.N. Sisakian, I.L. Solovtsov, and O. Shevchenko, 
Int. J. Mod. Phys. A {\bf 9}, 1929 (1994).

\bibitem{vpt6}  
W. Janke and H. Kleinert, 
Phys. Rev. Lett. {\bf 75}, 2787 (1995).

\bibitem{kneur}
J. -L. Kneur and M.B. Pinto,
Phys. Rev. Lett. {\bf 116}, 031601 (2016).
Phys.Rev. D {\bf 92}, 116008 (2015).

\bibitem{andersen1} 
J. O. Andersen, E. Braaten, and M. Strickland, 
Phys. Rev. Lett. {\bf 83}, 2139 (1999).


\bibitem{3loopglue1}N. Su., J. O. Andersen, and M. Strickland, 
Phys. Rev. Lett. {\bf 104}, 122003 (2010).
                    
\bibitem{3loopglue2} J. O. Andersen, M. Strickland, and N. Su, 
JHEP {\bf 1008}, 113 (2010).


          
\bibitem{3loopqcd1} 
J.O. Andersen, L.E. Leganger, M. Strickland and N. Su, 
Phys.\ Lett.\ B {\bf 696}, 468 (2011).
                    
\bibitem{3loopqcd2}
J.~O.~Andersen, L.~E.~Leganger, M.~Strickland and N.~Su, 
JHEP {\bf 1108}, 053 (2011).
                  
\bibitem{3loopqcd3}J.~O.~Andersen, L.~E.~Leganger, M.~Strickland and N.~Su, 
Phys.\ Rev.\ D {\bf 84}, 087703 (2011).

\bibitem{najmul3} 
N.~Haque, J.~O.~Andersen, M.~G.~Mustafa, M.~Strickland, and N.~Su,
Phys.\ Rev.\ D {\bf 89}, 061701 (2014).

\bibitem{complete}
N.~Haque, A.~Bandyopadhyay, 
J.~O.~Andersen, M.~G.~Mustafa, M.~Strickland, and N.~Su,
JHEP {\bf 1405}, 027 (2014).



\bibitem{purnendu1}  
P. Chakraborty, M. G. Mustafa, and M. H. Thoma, 
Eur. Phys. J. C. {\bf 23}, 591 (2002).

\bibitem{purnendu2}
P. Chakraborty, M. G. Mustafa, and M. H. Thoma, 
Phys. Rev. D {\bf 67}, 114004 (2003).

\bibitem{purnendu3}
P. Chakraborty, M. G. Mustafa, and M. H. Thoma, 
Phys. Rev. D {\bf 68}, 085012 (2003).

\bibitem{najmul11} 
N. Haque, M. G. Mustafa and M. H. Thoma, 
Phys. Rev. D {\bf 84}, 054009 (2011).

\bibitem{najmul12} 
N.~Haque and M.~G.~Mustafa,  
Nucl.\ Phys.\ A {\bf 862-863}, 271 (2011).

\bibitem{najmul13} 
N. Haque and M. G. Mustafa,
arXiv:1007.2076.

\bibitem{blaizotm3}
J.P. Blaizot, E. Iancu, and A. Rebhan,
Eur. Phys. J. C {\bf 27}, 433 (2003).

\bibitem{blaizotm2}
J.P. Blaizot, E. Iancu, and A. Rebhan 
Phys. Lett. B {\bf 523}, 143 (2001).

\bibitem{blaizotm1}
J.P. Blaizot, E. Iancu, and A. Rebhan,
Nucl. Phys. A {\bf 698}, 404 (2002).

\bibitem{blaizot1}
J.P. Blaizot, E. Iancu, and A. Rebhan,
Phys. Rev. Lett. {\bf 83}, 2906 (1999).

\bibitem{blaizot2}
J.P. Blaizot, E. Iancu, and A. Rebhan,
Phys. Lett. B {\bf 470}, 181 (1999).

\bibitem{blaizot3}
J.P. Blaizot, E. Iancu, and A. Rebhan,
Phys. Rev. D {\bf 63}, 065003 (2001).


\bibitem{twocolor1}
E. Dagotto, F. Karsch, and A. Moreo, 
Phys. Lett.  B {\bf 169}, 421 (1986); E. Dagotto, A. 
Moreo, and U. Wolff, Phys. Rev. Lett. {\bf 57}, 1292 (1986).

\bibitem{twocolor2}
S. Hands, J.B. Kogut, M.-P. Lombardo, S. E. Morrison, Nucl. Phys. B {\bf 558},
327 (1999).


\bibitem{alfiso} M. G. Alford, A. Kapustin, and
F. Wilczek, Phys.Rev. D {\bf 59}, 054502 (1999).


\bibitem{latt0} 	
J.~B. Kogut and D.~K. Sinclair,
Phys.\ Rev.\ D {\bf 66}, 014508 (2002).

\bibitem{latt1} 	
J.~B. Kogut and D.~K. Sinclair,
Phys.\ Rev.\ D {\bf 66},  034505 (2002).

\bibitem{latt2} 	
J.~B. Kogut and D.~K. Sinclair,
Phys.\ Rev.\ D {\bf 70}, 094501 (2004).

\bibitem{allton}
C.~R. Allton, M. Doring, S. Ejiri, S.~J. Hands, O. Kaczmarek, F. Karsch, E. 
Laermann, and K. Redlich,
Phys.\ Rev. \ D {\bf 71}, 054508 (2005).

\bibitem{ejiri}
S. Ejiri, F. Karsch, and K. Redlich,
Phys.\ Lett.\ B {\bf 633}, 275 (2006).

\bibitem{forc}
P. de Forcrand, M. A. Stephanov, and U. Wenger, 
PoS LAT2007, 237 (2007). 

\bibitem{detmold}
W. Detmold, K. Origonos, and Z. Shi, Phys. Rev. D {\bf 86}, 05407 (2012).

\bibitem{endiso}G. Endr\H{o}di, Phys. Rev. D {\bf 90}, 094501 (2014).
forc


\bibitem{stepson}
D. T. Son and M. A. Stephanov,
Phys. Atom. Nucl. {\bf 64}, 834 (2001).
\bibitem{borsanyi1}S.~Borsanyi, G.~Endrodi, Z.~Fodor, A.~Jakovac, S.~D.~Katz, 
S.~Krieg, C.~Ratti and K.~K.~Szabo, 
JHEP {\bf 1011}, 077 (2010).

\bibitem{borsanyi2} 
S.~Borsanyi, Z.~Fodor, S.~D.~Katz, S.~Krieg, C.~Ratti and K.~Szabo,
JHEP {\bf 1201}, 138 (2012)

\bibitem{Borsanyi:2012uq} 
S.~Borsanyi, S.~Durr, Z.~Fodor, C.~Hoelbling, S.~D.~Katz, S.~Krieg, D.~Nogradi 
and K.~K.~Szabo, B. C. Toth, and N. Trombitas,
JHEP {\bf 1208}, 126 (2012).
  
\bibitem{borsanyi3} Sz.~Bors\'anyi, G.~Endr\H{o}di, Z.~Fodor, S.D.~Katz, 
S.~Krieg, C.~Ratti and K.K.~Szab\'o,
JHEP {\bf 08}, 053 (2012).

\bibitem{borsanyi4} 
S.~Borsanyi,
Nucl.\ Phys.\ A904-905 {\bf 2013}, 270c (2013).

\bibitem{borsanyi5} 
S.~Borsanyi, Z.~Fodor, S.~D.~Katz, S.~Krieg, C.~Ratti and K.~K.~Szabo,
Phys.\ Rev.\ Lett.\  {\bf 111}, 062005 (2013).
  
\bibitem{sayantan}
S.~Sharma,
Adv.\ High Energy Phys.\  {\bf 2013}, 452978 (2013).

\bibitem{bnlb0} 
F.~Karsch, B.~-J.~Schaefer, M.~Wagner and J.~Wambach,
Phys.\ Lett.\ B {\bf 698}, 256 (2011).

\bibitem{bnlb1} 
A.~Bazavov, H.~-T.~Ding, P.~Hegde, O.~Kaczmarek, F.~Karsch, E.~Laermann, 
Y.~Maezawa and S.~Mukherjee \emph{ et al.},
Phys.\ Rev.\ Lett.\  {\bf 111}, 082301 (2013).

\bibitem{bnlb2} 
A.~Bazavov, H.~-T.~Ding, P.~Hegde, F.~Karsch, C.~Miao, S.~Mukherjee, 
P.~Petreczky and C.~Schmidt, and A. Velytsky,
Phys. Rev. D {\bf 88}, 094021 (2013). 

\bibitem{bnlb3} 
A.~Bazavov, H.~T.~Ding, P.~Hegde, O.~Kaczmarek, F.~Karsch, E.~Laermann, S.~Mukherjee and P.~Petreczky \emph{ et al.},
Phys.\ Rev.\ Lett.\  {\bf 109}, 192302 (2012).
  
\bibitem{milc} 
C.~Bernard \emph{ et al.},
Phys.\ Rev.\ D {\bf 71}, 034504 (2005).

\bibitem{hotqcd1} 
A.~Bazavov, T.~Bhattacharya, M.~Cheng, N.~H.~Christ, C.~DeTar, S.~Ejiri, 
S.~Gottlieb and R.~Gupta \emph{ et al.},
Phys.\ Rev.\ D {\bf 80}, 014504 (2009).

\bibitem{hotqcd2} 
A.~Bazavov \emph{ et al.},
Phys.\ Rev.\ D {\bf 86}, 034509 (2012).

\bibitem{peter_review} 
P.~Petreczky,
J.\ Phys.\ G {\bf 39}, 093002 (2012).

\bibitem{hotqcd15}
HotQCD Collaboration (A. Bazavov (Iowa U.) et al.),
Phys.\ Rev.\ D {\bf 90}, 094503, (2014).

\bibitem{dingrev}
H.-T. Ding, F. Karsch, S. Mukherjee,
Int. J. Mod. Phys. E {\bf 24}, 1530007 (2015).
\bibitem{bellwied} 	
R. Bellwied, S. Borsanyi, Z. Fodor, S.D. Katz, A. Pasztor, C. Ratti, 
and K.~K. Szabo,
Phys. Rev. D {\bf 92}, 114505 (2015).

\bibitem{ding}
H. -T. Ding, S. Mukherjee, H. Ohno, P. Petreczky, and H. -P. Schadler,
Phys. Rev. D {\bf 92}, 074043 (2015).


\bibitem{lagrangian} 
E.~Braaten and R.~D.~Pisarski,
Phys.\ Rev.\ D {\bf 45}, R1827 (1992).



\bibitem{braatennieto2} 
E.~Braaten and A.~Nieto,
Phys.\ Rev.\ D {\bf 53}, 3421 (1996).

\bibitem{run1}
D. J. Gross and F. Wilczek,
Phys. Rev. Lett. {\bf 30}, 1343 (1973).
\bibitem{run2}
H. D. Politzer,
Phys. Rev. Lett. {\bf 30}, 1346 (1973).

\bibitem{latticealpha}
A.~ Bazavov, N.~Brambilla, X.~Garcia i Tormo, P.~Petreczky, J.~Soto and 
A.~Vairo,
Phys.\ Rev.\ D {\bf 86}, 114031 (2012).




\end{thebibliography}
\bibliographystyle{apsrmp4-1}

\end{document}